# Fourier transform spectroscopy around 3 μm with a broad difference frequency comb

Samuel A. Meek [1], Antonin Poisson [1,2], Guy Guelachvili [2], Theodor W. Hänsch [1,3], Nathalie Picqué [1,2,3] *

1. Max Planck Institut für Quantenoptik, Hans-Kopfermann-Str. 1, 85748 Garching, Germany
2. Institut des Sciences Moléculaires d'Orsay, CNRS, Bâtiment 350, Université Paris-Sud, 91405 Orsay, France
3. Ludwig-Maximilians-Universität München, Fakultät für Physik, Schellingstrasse 4/III, 80799 München, Germany
* corresponding author: nathalie.picque@mpq.mpg.de

*Abstract*
*We characterize a new mid-infrared frequency comb generator based on difference frequency generation around 3.2 μm. High power per comb mode (>$10^{-7}$ W/mode) is obtained over a broad spectral span (>700 nm). The source is used for direct absorption spectroscopy with a Michelson-based Fourier transform interferometer.*

Lasers frequency combs [1] are opening up new opportunities for broad spectral bandwidth direct absorption spectroscopy. In recent times, a variety of novel techniques have demonstrated improved capabilities in terms of sensitivity, acquisition times, resolution and/or accuracy. Such promising experiments have been initially developed [2-5] in the near-infrared region, where ultra-short pulse lasers are conveniently available. For spectroscopic applications, the mid-infrared spectral region (2-20 μm) is more appealing because most molecules have strong and characteristic fundamental ro-vibrational transitions that provide a "fingerprint" of the molecule. As the technology in this region is technically more demanding, considerable efforts have been undertaken in recent years to develop new frequency comb generators based on lasers directly emitting in the mid-infrared [6], nonlinear frequency conversion either by difference frequency generation [7-9] or optical parametric oscillation [10,11], or Kerr nonlinearity in high-quality factor whispering-gallery mode microresonators pumped by a continuous-wave laser [12]. Concurrently, a number of spectrometric techniques have been explored [13-15] to efficiently analyze the comb light. Mid-infrared frequency comb sources and their applications have been discussed in a recent review [16].

In this letter, we report on the applicability of a new high-power mid-infrared frequency comb to direct absorption frequency comb spectroscopy. A schematic of this source is shown in Fig. 1. An erbium-doped fiber oscillator with a repetition frequency of 100 MHz emits around 1.55 μm and seeds two erbium-doped fiber amplifiers. The output of one of those is launched into a nonlinear fiber that shifts the spectrum to 1.04 μm where it is reamplified in two steps with an ytterbium preamplifier and an ytterbium power amplifier. The outputs of the erbium (signal, 100fs, 400 mW) and ytterbium (pump, 100 fs, 1.4W) amplifiers are combined on a dichroic mirror and the temporal superposition of the pulses is achieved by introducing extra optical delay into the path of the erbium amplifier. The beams are focused onto a 3-mm long periodically-poled MgO-doped lithium niobate crystal with poling period varying from 27 to 32 μm. The crystal temperature is stabilized to 92°C. The generated mid-infrared idler beam is isolated by means of a wedged germanium filter. The mid-infrared beam has an average power of 120 mW and a peak spectral power density of about 0.32 mW/nm. Fig. 2a shows the idler's autocorrelation trace, which has a full width at half maximum of 150 fs, corresponding to a pulse duration of 106 fs for a Gaussian pulse shape. The shape of the autocorrelation trace, however, indicates some spectral chirp, consistent with the width of the associated spectrum (Fig. 2b). The spectrum (Fig 2b) is centered at 3.1 µm, with a spectral span from 2.7 to 3.4 µm. The lines present in the spectrum are due to atmospheric water vapor. Slight tunability from 2.3 to 3.6 µm is obtained by adjusting the optical delay between pump and signal.

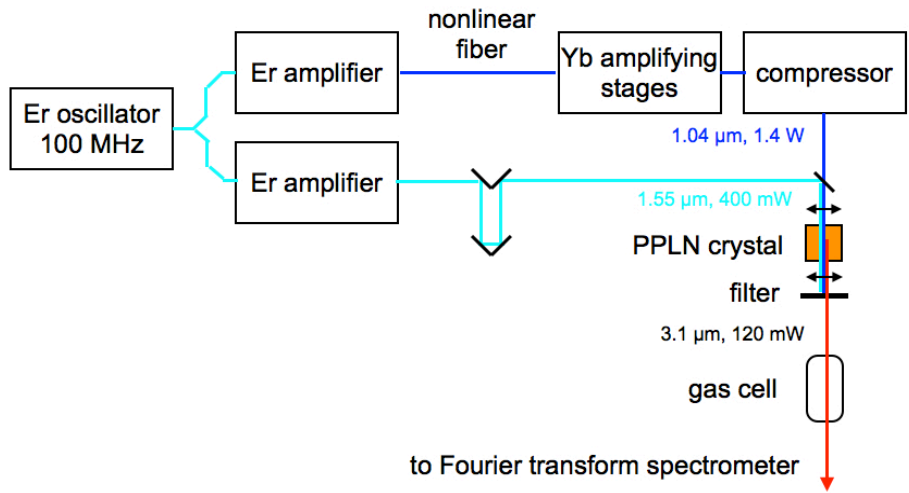

*Figure 1: Experimental set-up used to generate femtosecond mid-infrared pulses and perform absorption spectroscopy. A femtosecond erbium-doped oscillator is used to generate the 1.04-µm pump (through spectral shift in a nonlinear fiber and amplification) and the 1.55-µm signal beams. The two beams are mixed in a periodically-poled lithium niobate crystal. The generated 3.1-µm mid-infrared beam is filtered. It interrogates a gas cell and it is analyzed by a Fourier transform spectrometer.*

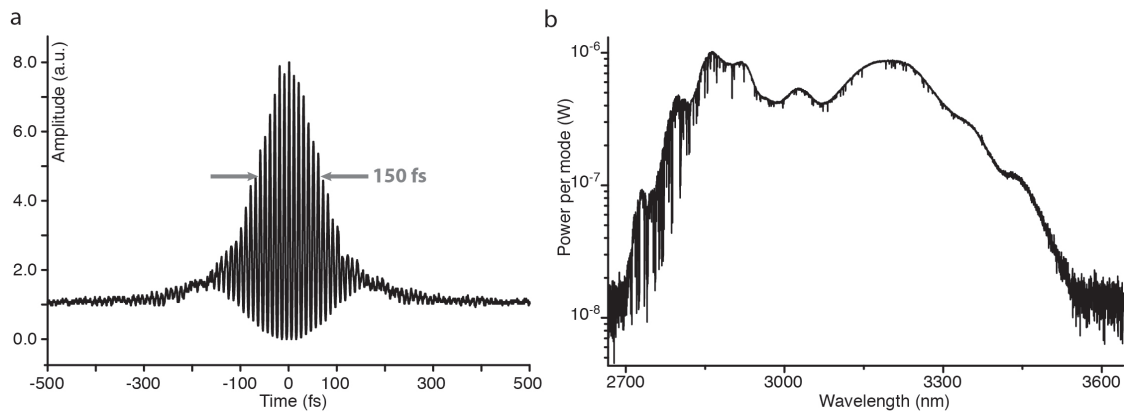

*Figure 2: (a) Interferometric autocorrelation of the pulses of the idler of the difference-frequency generation source. (b) Spectrum of the idler radiation, measured with a Fourier transform spectrometer at 6 GHz resolution.*

The heterodyne beat between the idler and a continuous-wave optical parametric oscillator (OPO) tunable between 2.4 to 3.2 μm (specified linewidth <1 MHz at 1 s) is measured with a InGaAsSb photodetector that has a 100-MHz bandwidth. The result suggests good pulse-to-pulse coherence. Figure 3a shows the result of the measurement of the free-running beat note at the wavelength of 3.0 μm with an integration bandwidth of 300 kHz, chosen to limit the spread in beat-note frequency due to variation in the unlocked optical parametric oscillator source.

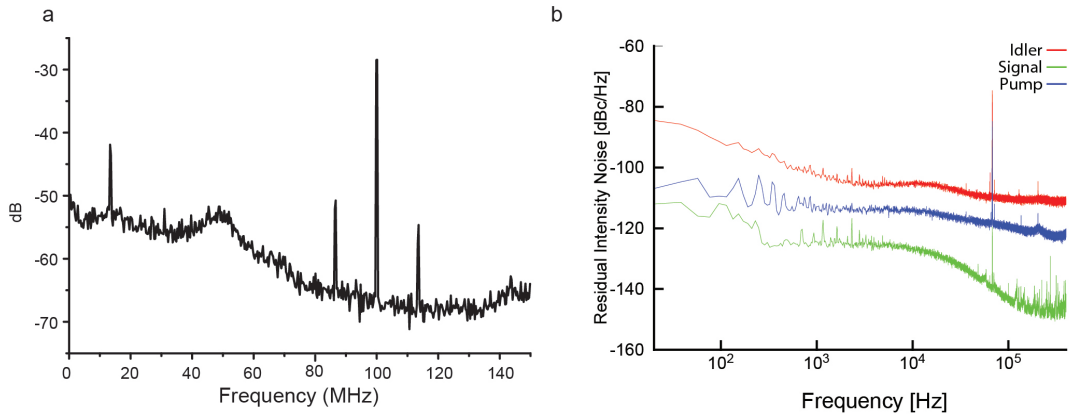

*Figure 3: (a) Free-running beat note between the mid-infrared idler frequency comb and a continuous-wave OPO. The 100 MHz line is the repetition frequency of the comb and the other lines are the beat notes of comb lines with the OPO. (b) Relative intensity noise of the idler, signal and pump from 50 to $5\ 10^5$ Hz.*

To quantify the magnitude of noise sources, relative intensity noise (RIN) measurements at each stage of the system are performed by monitoring (Fig. 3b) each intermediate output with a fast photodiode and using a signal analyzer to obtain a power spectral density of the relative intensity noise. The pump beam is the noisiest part of the system and we infer that the main source of RIN originates in the nonlinear fiber. The RIN is almost flat from 1 to 30 MHz with values of the order of -149, -124 and -119 dBc/Hz for the signal, pump and idler, respectively. There is a noise increase of more than 30 dB in this frequency range between the erbium oscillator and the idler output. This value is achieved after careful optimization of the temporal overlap between pump and signal pulses and can easily reach higher values. Interferometers, like Michelson or dual-comb spectrometers, however, allow for efficient amplitude noise cancellation when the two outputs of the interferometer are subtracted. This seems an important prerequisite for efficient spectroscopic measurement with this light source.

To demonstrate the potential for spectroscopy, we insert (Fig. 1) a 5-cm long single pass cell filled with 67 hPa of acetylene in natural abundance in the idler's beam path and we analyze the light transmitted by the cell with a Michelson-based Fourier transform spectrometer. The beam is attenuated before being focused onto the J-stop of the interferometer. Fig. 4 shows a spectrum at 6-GHz resolution over the full spectral span, while the inset zooms onto some of the acetylene lines and gives their assignment [17]. The crowded spectrum is mainly composed of the lines of the $\nu_3$ and $\nu_2+\nu_4+\nu_5$ cold bands of $^{12}C_2H_2$, with rovibrational levels interacting in a Fermi-type resonance. The resolution is currently limited by the Fourier transform spectrometer, but the repetition frequency of the comb, 100 MHz, allows for Doppler-limited measurements. The frequency comb generator used as a light source for absorption Fourier transform spectroscopy features the same advantages as those already pointed out in other publications, e.g. [6], but has the additional interest that it enables a very broad spectral span in the region of the CH, OH, NH stretches in molecules.

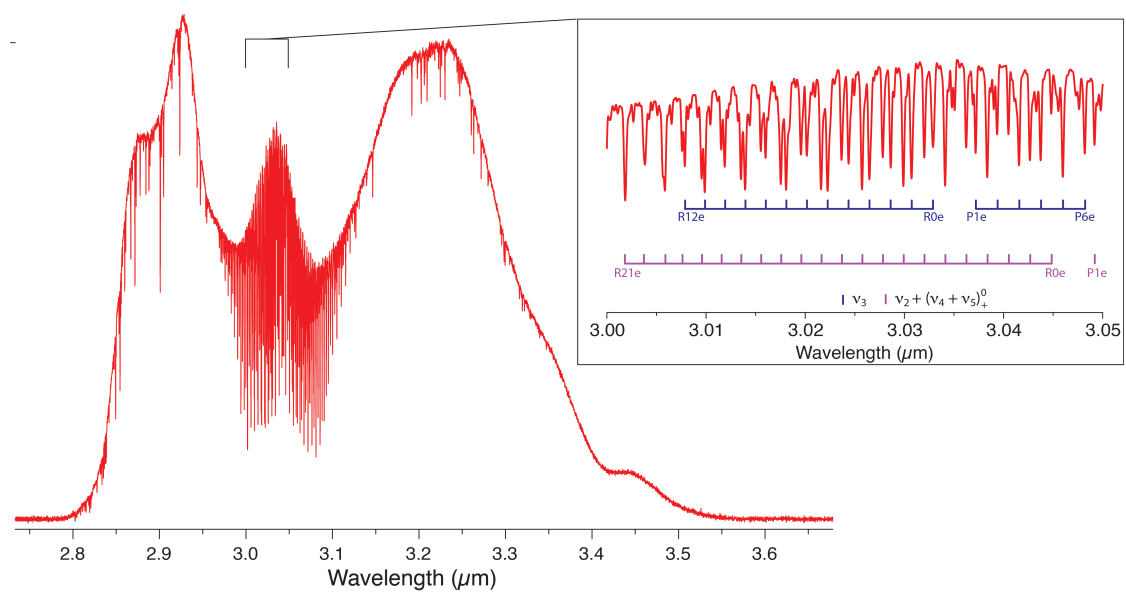

*Figure 4: Spectrum of $C_2H_2$ in the 3.0 μm region illustrating the spectral bandwidth capabilities of the spectrometric technique. The $\nu_3$ and $\nu_2+\nu_4+\nu_5$ bands are observed. The left graph shows the entire spectral domain covered in a single recording while the right inset shows a portion of the spectrum expanded in wavelength and intensity scales.*

Figure 5 summarizes the main characteristics of the new frequency comb generator and compares them to a non-exhaustive selection of the state-of-the art recent reports. The system achieves very high power per comb line (>$10^{-7}$ W/mode over a 730 nm span) when compared to other frequency comb sources based on difference frequency generation. Thus the difference frequency comb generator reaches a domain that has traditionally been occupied by optical parametric oscillators. Although noisier than an OPO, a fiber-based difference frequency generation source is easier to operate, as it is mostly alignment free. Interestingly, the generation of difference frequencies between the different teeth of a single comb oscillator leads to comb lines with a frequency that is an integer multiple of the repetition frequency. The carrier–envelope offset frequency is indeed fixed to zero. In view of dual-comb spectroscopy applications, cancellation of the carrier-envelope offset frequency is expected to significantly simplify experimental implementation, either with stabilized [3,14] or free-running [18] set-ups. Furthermore, on the basis of the recent demonstrations of nonlinear dual-comb spectroscopy [19-21] in the near-infrared region, the high power per comb mode makes it possible to envision mid-infrared nonlinear dual-comb spectroscopy, e.g. sum-frequency generation at surfaces or two-photon excitation of rovibrational molecular transitions.

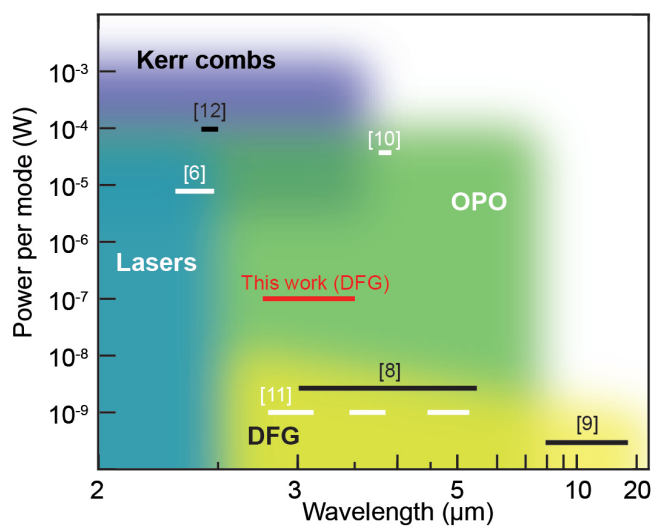

*Figure 5: Typical spectral regions and power per comb mode reached with mid-infrared femtosecond lasers (blue), difference frequency generation DFG (yellow), optical parametric oscillators OPOs (green) and microresonator-based Kerr combs (violet). Performances of a few selected sources are represented with black and white lines, with citations shown in brackets. The width of each bar represents the spectral span (regardless of their tunability) over which the respective comb has at least the power per mode indicated by its position on the ordinate. The characteristics of the source investigated in the present work are displayed in red. This figure is adapted from a figure displayed in [16].*

**Acknowledgments**
The frequency comb source has been developed in collaboration with Menlo Systems GmbH within a Eurostars project. Research conducted in the scope of the European Laboratory for Frequency Comb Spectroscopy. Support by the Max Planck Foundation and the Munich Center for Advanced Photonics also are acknowledged.